\documentstyle[12pt]{article}
\begin{document}
\rightline{hep-th/9406188, CGPG-94/6-2, VPI-IHEP-94-5}
\vskip .2in
\centerline{\Large The Standard Model with Gravity Couplings}
\vskip .2in
\vskip 0.2in
\centerline{Lay Nam Chang ${}^*$}
\centerline{Department of Physics,}
\centerline{Virginia Tech,}
\centerline{Blacksburg, VA 24061-0435, U.S.A.}
\vskip 0.10in
\centerline{and}
\vskip 0.10in
\centerline{Chopin Soo ${}^\dagger$}
\centerline{Center for Gravitational Physics and Geometry,}
\centerline{Department of Physics,}
\centerline{Pennsylvania State University,}
\centerline{University Park, PA 16802-6300, U.S.A.}
\vskip 0.20in
\centerline{PACS number(s): 04.60.-m, 04.62.+v, 11.15.-q, 11.30.Er}
\vskip 0.20in

In this paper, we examine the coupling of matter fields to gravity within
the framework of the Standard Model of particle physics.   The coupling is
described
in terms of Weyl fermions of a definite chirality, and employs
only (anti)self-dual or left-handed spin connection fields.
It is known from the work of Ashtekar and others that such fields
can furnish a complete description of gravity without matter.   We
show that conditions
ensuring the cancellation of perturbative chiral
gauge anomalies are not disturbed.  We also explore a global
anomaly associated with the theory, and
argue that its removal requires that the number of fundamental
fermions in the theory must be multiples of 16.
In addition, we investigate the behavior of the theory under discrete
transformations P, C and T; and discuss possible violations of these
discrete symmetries, including CPT, in the presence of instantons and the
Adler-Bell-Jackiw anomaly.

\vfill
$^*$Electronic address: laynam@vtvm1.cc.vt.edu \hfil
\vskip 0.0in
$^\dagger$Present address:Dept. of Physics, Virginia Tech, Blacksburg,
VA 24061-0435; electronic address: soo@vpihe1.phys.vt.edu \hfil

%
%
\section*{I. Introduction}
\bigskip

Ashtekar \cite{ash} has introduced a set of variables to describe gravity,
which makes essential use of the chiral decomposition of the connection
one-forms of the local Lorentz group. What has been shown is that, at least
for Einstein manifolds without matter, the full set of field
equations of General Relativity can be recovered by use of only one
of the two chiral projections of these connection forms.   We may
either use the self-dual or the anti-self-dual connections and their
respective conjugate variables. The constraints and reality conditions of
the theory then define what the other set has to be.

Chiral projections are used routinely in particle physics; indeed,
the fermion fields that define the Standard Model are all chiral
Weyl spinor fields.    Within the Ashtekar context, left-handed
spinor fields are coupled to one of these connection forms, say
$A^-$, while right-handed ones are coupled to the other.
Since only one of these is all one needs to define general
relativity, the actual Lagrangian must be expressed entirely in
terms of either left-handed or right-handed Weyl spinor fields.

That the Standard Model can be so described is of course already known.
For example, in $SO(10)$ grand unification schemes \cite{gut},
one employs a single 16-dimensional left-handed Weyl field to describe one
generation of fermions.  What is less clear is how the coupling
of Ashtekar fields affects the resultant physics.

In what follows, we describe some facets of these consequences.
Since Ashtekar gravity makes use of only one of the chirally
projected Lorentz connections, there arises the question of
whether anomalies which are normally present in such theories are
under control.
We show that the usual conditions for anomaly cancellations for the
Standard Model gauge groups remain true in the presence of Ashtekar
gravity, and that the new fields do not introduce any further
perturbative anomalies.    However, they do introduce global
anomalies.   Cancellation of these obstructions for Grand Unified Theories
in the most general context results in the
rather strong constraint that the total number of fundamental
fermions in the theory must be
a multiple of 16.  Grand unification schemes based upon groups such
as $SU(5)$ are therefore inconsistent when coupled to gravity.
As a consequence, there is every likelihood that neutrinos must be
massive when consistency with gravity is taken into account.
In particular, the $SO(10)$ GUT with 16 fundamental Weyl fermions per
generation is singled out as the preeminent and simplest choice free from
the global anomaly, if one allows for generalized spin structures and
arbitrary topologies.

We also discuss how the usual discrete symmetries
are implemented in the presence of
Ashtekar gravity.
We shall show that it is possible to
posit discrete transformation
laws for the fermion and Ashtekar fields which are consistent both with
our general notions of what parity, time-reversal, and charge conjugation
transformations are, and with the fundamental canonical commutation
relations of all of the fields. Discrete symmetries for bispinors can be
implemented in the classical limit modulo certain reality conditions.
However, what we shall
show is that, inevitably,
the underlying quantum theory is not invariant under
parity due to the occurrence axial anomaly in quantum field theory.
The question of CPT invariance will also be briefly discussed.

\bigskip
\section*{II. The Samuel-Jacobson-Smolin action and the Ashtekar variables}
\bigskip
We first consider spacetimes of Lorentzian signature $(-, +, +, +)$
and start with the gravitational action proposed by Samuel, and Jacobson and
Smolin \cite{samuel}
\begin{eqnarray}
{\cal S^{\mp}_{SJS}}&=& \int_{M} L^{\mp}_{G}
\nonumber\\
&=&{1\over {8 \pi G}}\int_{M}\Sigma^{\mp{a}}\wedge F^{\mp}_{a}\quad
{\pm}\quad
i{\lambda \over 3(16\pi{G})}\int_{M}\Sigma^{\mp{a}}\wedge
\Sigma^{\mp}_a
\end{eqnarray}
The (anti)self-dual two-forms $\Sigma^{\mp}$ which obey
$\ast \Sigma^{{\mp}a} = {\mp}i\Sigma^{{\mp}a}$, are defined as
\begin{equation}
\Sigma^{{\mp}a} \equiv (-e^{0}\wedge e^{a} \pm {i \over 2}
\epsilon^a\,_{bc}e^b \wedge e^c).
\end{equation}
$F^{\mp}$ are the curvature two-forms of the $SO(3, C)$ Ashtekar
connections i.e.
\begin{equation}
F^{\mp}_{a} = dA^{\mp}_{a} + {1\over 2}\epsilon_{a}\,^{bc}A^{\mp}_{b}
\wedge A^{\mp}_{c},
\end{equation}
and $e_A,A=0,...,3$ denote the vierbein one-forms in four dimensions;
$\epsilon_{abc} \equiv \epsilon_{0abc}$ while $\lambda$ is the
cosmological constant. Latin indices label
flat Lorentz indices while spacetime indices will be denoted by Greek indices.
Lower case Latin indices run from 1 to 3 while upper case Latin
indices range from 0 to 3.

The Ashtekar variables \cite{ash} are sometimes referred to as
(anti)self-dual variables
because the equations of motion of the Samuel-Jacobson-Smolin action with
respect to $A^{\mp}$,
\begin{equation}
D^{\mp}\Sigma^{\mp{a}} = 0,
\end{equation}
imply that $A^-$ and $A^+$ are the anti-self-dual and
self-dual part of
the spin connection, $\omega$, respectively i.e.
\begin{equation}
A^{\mp}_a = {\pm}i\omega_{oa} - {1\over 2}\epsilon_a\,^{bc}\omega_{bc}.
\end{equation}

It is easy to see that the action reproduces the Ashtekar variables and
constraints. For convenience, we work in the spatial gauge in which the
components of the vierbein and its inverse can be written in the form
\begin{equation}
e_{A\mu}=\left[\matrix{ N &0 \cr N^je_{aj}& e_{ai}}\right]\quad,\quad
E^{\mu}\,_{A}=\left[\matrix{N^{-1}& {0}\cr
-{(N^i/N)}&{\sigma^i\,_a}}
\right].
\end{equation}
The form assumed in (6) is compatible with the ADM \cite{adm} decomposition
of the metric
\begin{eqnarray}
ds^{2}&=& e_{A\mu}{e^A}_\nu dx^\mu dx^\nu
\nonumber\\
&=& -N^2(dx^0)^2+g_{ij}(dx^i+N^idx^0)(dx^j+N^jdx^0)
\end{eqnarray}
with the spatial metric $g_{ij}=e^{a}\,_{i}e_{aj}$. Thus we see that the
choice (6) in no way compromises the values of the lapse and shift
functions, $N$ and $N^i$, which have geometrical interpretations in
hypersurface deformations. With this decomposition, it is
straightforward to rewrite
\begin{eqnarray}
{\cal S}^{\mp}_{SJS}&=&
{1 \over 16\pi{G}}
\int d^4x \left\{{\pm}2i\tilde\sigma^{ia}\dot A^{\mp}_{ia}\quad
{\pm}\quad 2i A^{\mp}_{0a} D_i\tilde\sigma^{ia} \quad{\pm}\quad 2i
N^j\tilde\sigma^{ia}
F^{\mp}_{ija}\right\}
\nonumber\\
&&-{1 \over 16\pi{G}}
\int d^4x \left\{{{\rlap{\lower2ex\hbox{$\,\,\tilde{}$}}{N}}}
\left(\epsilon_{abc}
\tilde\sigma^{ia}\tilde\sigma^{jb}F^{{\mp}c}_{ij}+{\lambda\over 3}
\epsilon_{abc}
{{\rlap{\lower2ex\hbox{$\,\,\tilde{}$}}{\epsilon_{ijk}}}}
\tilde\sigma^{ia}\tilde\sigma^{jb}\tilde\sigma^{kc}\right)\right\}
\nonumber\\
&& +  {\rm boundary \, terms},
\end{eqnarray}
with $\tilde\sigma$ and ${\rlap{\lower2ex\hbox{$\,\,\tilde{}$}}{N}}$
defined as
\begin{eqnarray}
\tilde\sigma^{ia}&\equiv& {1\over 2} \tilde
\epsilon^{ijk}\epsilon^{abc}e_{jb}e_{kc}
\nonumber\\
{\rlap{\lower2ex\hbox{$\,\,\tilde{}$}}{N}} &\equiv& \det (e_{ai})^{-1}N .
\end{eqnarray}
The tildes above and below the variables indicate that they are tensor
densities of weight $1$ and $-1$ respectively. Therefore,
${\pm}(2i\tilde\sigma^{ia}/16{\pi}G)$ are
readily identified as the conjugate variables to $A^{\mp}_{ia}$, and we
have the commutation relations
\begin{equation}
\left[{\tilde\sigma}^{ia}({\vec x},t),A^{\mp}_{jb}({\vec y},t)\right]
= {\pm}(8\pi{G})\delta^i_j \delta^a_b \delta^3({\vec x} - {\vec y}).
\end{equation}

The variables $A^{\mp}_{0a}$, $N^i$ and
${\rlap{\lower2ex\hbox{$\,\,\tilde{}$}}{N}}$ are clearly Lagrange
multipliers for the Ashtekar constraints, which can be identified as
Gauss' law generating $SO(3)$ gauge invariance
\begin{equation}
G^a\equiv 2iD_i\tilde\sigma^{ia}\approx 0,
\end{equation}
and the supermomentum and ``superhamiltonian constraints
\begin{eqnarray}
H_i &\equiv&
2i\tilde\sigma^{ja}F_{ija}\approx 0
\nonumber\\
H&\equiv&
\epsilon_{abc}\tilde\sigma^{ia}\tilde\sigma^{jb}(F^c_{ij}+
{\lambda\over3}
{{\rlap{\lower2ex\hbox{$\,\,\tilde{}$}}{\epsilon_{ijk}}}}
\tilde\sigma^{kc})\approx 0
\end{eqnarray}
respectively.
Ashtekar showed that these constraints and their algebra, despite their
remarkable simplicity, are equivalent in content to the constraints and
constraint
algebra of general relativity \cite{ash}. Note
that both the self-dual and anti-self-dual versions,
${\cal S^{\pm}_{SJS}}$, describe {\it pure}
gravity equally well.
%

\bigskip
\section*{III. Coupling to matter fields}
\bigskip

The coupling of matter fields to gravity
described by the (anti)self-dual Ashtekar variables have been considered by
others before \cite{ART, Ja, Kod}, and will be examined closely in this work.
It should be emphasized that besides self-interactions in pure gravity, only
fermions couple directly to the Ashtekar connections. They are hence
direct sources for the Ashtekar connection. Conventional scalars and
Yang-Mills fields have direct couplings
only to the metric (hence to $\tilde \sigma$) rather than
to the Ashtekar-Sen connection. Thus when one substitutes the
conventional gravitational
action with the Samuel-Jacobson-Smolin action, complications can come
from the fermionic sector, primarily because one has to make a choice
between the actions described by $A^+$ and $A^-$ and once the choice is
made, it is permissible to couple only right or left-handed fermions (but
not both) to the theory.\footnote{We shall choose the $-$ action.
There is an ambiguity in the conventions of the Dirac matrices, $\gamma^{A}$,
which allows one to couple
either $A^+$ or $A^-$ to left-handed Weyl spinors. We have adopted a
choice which couples anti-self-dual spin connections and $A^-$ to
left-handed spinors.
To the extent that there are no right-handed neutrinos in nature, we
should describe nature with only left-handed Weyl spinors. So once the
initial choice of coupling $A^-$ to the neutrino is made, the theory
cannot be described by $A^+$ and {\it right-handed} Weyl spinors.}

We first look at the conventional Dirac action,  with couplings
to ordinary spin connections. We then describe
couplings of Ashtekar fields to fermions, and explore the physical
implications in the following sections.

Consider the conventional Dirac action for
an electron or a single quark of a particular color in the presence of
only the conventional gravitational field. The action can be written as
\begin{equation}
{\cal S}_{D} = -{i \over 2}\int_{M}(\ast 1){\overline \Psi}
\gamma^{A}{{E_{A}}_\lfloor}D_{\omega}\Psi + {h. c.}
\end{equation}
where the covariant derivative with respect to the spin connection,
$\omega$, is defined by
\begin{equation}
D_{\omega}\Psi = dx^{\mu}(\partial_{\mu} + {1\over 2}{\omega_{{\mu}BC}}
{\cal S}^{BC})\Psi
\end{equation}
with the generator ${\cal S}^{AB} = {1\over 4}[\gamma^{A},\gamma^{B}]$.
We adopt the convention
\begin{equation}
\{ \gamma^{A}, \gamma^{B}\}= 2\eta^{AB}
\end{equation}
with $\eta^{AB} = {\rm diag}(-1,+1,+1,+1)$.

We use the chiral
representation henceforth for convenience and clarity. In the chiral
representation
\begin{equation}
\gamma^{A} = \left(\matrix{ 0& i\tau^{A}\cr i{\overline \tau}^{A}&0\cr}
\right),
\end{equation}
where $\tau^{a}= -{\overline \tau}^{a}$ ($a$=1,2,3) are Pauli matrices, and
$\tau^{0} = {\overline \tau}^{0} = -I_{2}$. We also have
\begin{eqnarray}
\gamma^{5}&=& {i}\gamma^0\gamma^1\gamma^2\gamma^3
\nonumber\\
&=& \left(\matrix{I_{2}&0\cr 0&-I_{2}}\right)
\end{eqnarray}
and the Dirac bispinor is expressed in terms of two-component left and
right-handed Weyl spinors, $\phi_{L,R}$, as
\begin{equation}
\Psi = \left(\matrix{ \phi_{R}\cr \phi_{L} \cr}\right) .
\end{equation}
The contractions in (13) are defined by
\begin{equation}
{{E_{A}}_\lfloor}D\Psi \equiv E^{\mu}\,_{A}{\cal D}_{\mu}\Psi .
\end{equation}
Note that the vierbein vector fields and one-forms, $E_{A} = E^{\mu}\,_{A}
\partial_\mu$ and $ e^{A}= e^{A}\,_{\mu}dx^{\mu}$,
satisfy ${E_{A}}\lfloor{e^{B}} = \delta_{A}\,^{B}$.

In the chiral representation, the covariant derivative can be written as
\begin{equation}
D\Psi \equiv dx^\mu {\left\{{\partial_\mu}I_{4}
-{i}\left(\matrix{A^{+}_{\mu a}{\tau^a \over 2} &0 \cr
0 & A^{-}_{\mu a}{\tau^a \over 2}}\right)\right\}}{\left(\matrix{\phi_{R}\cr
\phi_{L}}\right)} ,
\end{equation}
and in conventional coupling of fermions to spin connections, $A^{\mp} $ are
precisely
\begin{equation}
{A^{\mp}_a} = \pm {i}{\omega}_{0a}-{ 1\over 2}\epsilon_{a}\,^{bc}\omega_{bc}
\end{equation}

Written in terms of the Weyl spinors, the Dirac action is
\begin{eqnarray}
{\cal S}_{D}&=&
\overbrace{\int_{M}(\ast 1)(-{i\over 2})
\phi^\dagger_{L}\tau^{A}{E_{A}}_\lfloor
{D^{-}}\phi_{L}}^{(1)}+\overbrace{\int_{M}(\ast 1)({i\over 2})
{E_{A}}_\lfloor(D^{-}\phi_{L})^\dagger \tau^{A}\phi_{L}}^{(2)}
\nonumber\\
&+&\underbrace{\int_{M}(\ast 1)(-{i\over 2})
\phi^\dagger_{R}{\overline\tau}^A{E_{A}}_\lfloor{D^{+}}\phi_{R}}_{(3)}
+\underbrace{\int_{M}(\ast 1)({i\over 2})
{E_{A}}_\lfloor(D^{+}\phi_{R})^\dagger
{\overline\tau}^{A}\phi_{R}}_{(4)}
\end{eqnarray}
where
\begin{eqnarray}
{D^-}\phi_{L}\equiv (d - {i}A^{-}_{a}{\tau^a \over 2})\phi_{L}\quad &,& \quad
{D^+}\phi_{R}\equiv (d - {i}A^{+}_{a}{\tau^a \over 2})\phi_{R}
\nonumber\\
({{D^-}\phi_{L}})^{\dagger}\equiv d{\phi_{L}}^\dagger +
{i}{\phi_{L}}^\dagger{A^{+}_{a}}{\tau^a \over 2} \quad &,& \quad
({{D^+}\phi_{R}})^\dagger\equiv d{\phi_{R}}^\dagger
+ {i}{\phi_{R}}^\dagger{A^{-}_{a}}{\tau^a \over 2}
\end{eqnarray}
Notice that in (22), terms (1) and (4) contain $A^{-}$ but not $A^{+}$,
while (2) and (3) involve $A^{+}$ but not $A^{-}$. Furthermore,
it is well-known that right-handed spinors can be
written in terms of left-handed ones through the relation
\begin{equation}
\phi_R = -i\tau^2 \chi^{\ast}_L
\end{equation}
and vice-versa,
so the term (4) which involves $A^-$ can be written in terms of totally
{\it left-handed} (anti-self-dual) spin connections and Weyl spinors as
\begin{equation}
\int_M (\ast 1)(-{i \over 2}){ \chi^\dagger_{L}}\tau^{A}{E_{A}}_\lfloor{D^{-}}
\chi_{L}
\end{equation}
Similar remarks apply to the term (2) with regard to right-handed spinors
and the field $A^+$.

In order to couple spinors to Ashtekar connections of only one chirality,
we can now define modified ``chiral" Dirac actions
\begin{eqnarray}
{\cal S}^{-}_{D}&\equiv & 2\{(1) + (4)\}
\nonumber\\
&=&\int_{M}(\ast 1)(-i)({\phi^\dagger_{L}\tau^{A}{E_{A}}_\lfloor
{D^{-}}\phi_{L}
+ \chi^\dagger_{L}\tau^{A}{E_{A}}_\lfloor{D^{-}}\chi_{L}})
\end{eqnarray}
\begin{eqnarray}
{\cal S}^{+}_{D}&\equiv& 2\{(2) + (3)\}
\nonumber\\
&=&\int_{M}(\ast 1)(-i)({\phi^\dagger_{R}{\overline\tau^{A}}{E_{A}}_\lfloor
{D^{+}}\phi_{R}+{{E_{A}}_\lfloor}{(D^{-}\phi_{L})}^\dagger\tau^{A}\phi_{L})} .
\end{eqnarray}
Next, we make the identification that the quantities $A^{\mp}$, as
anticipated by the notation, are precisely
the connections introduced by Ashtekar in his simplification of the
constraints of General Relativity, and the very same variables in the
Samuel-Jacobson-Smolin action of Eq. (1). Instead of the conventional
sum of the Einstein-Hilbert-Palatini action and the full Dirac action,
${\cal S_D}$, the total action is now taken to be
($\cal S^-_D + \cal S^-_{SJS}$).
Remarkably, it is possible to show (for instance, by using Eq. (37)) that
this total action reproduces the correct classical equations of motion for
General Relativity with spinors \cite{ART, rema}.

\bigskip
%
%
\section*{IV. Discrete transformations and spinors}
\bigskip

The usual discrete transformations for the second quantized spinor fields
are
\begin{eqnarray}
P &:& {\phi_L}(x) \longleftrightarrow  -i\tau^2\chi_L^{\ast}(P^{-1}(x))
\qquad {\rm i.e.}\quad (\phi_L(x) \longleftrightarrow \phi_R(P^{-1}(x)),
\nonumber\\
C &:& {\phi_L}(x)\longleftrightarrow \chi_L(x)
\qquad {\rm i.e.} \quad (\Psi^c = C{\overline \Psi}^T ),
\nonumber\\
T &:& {\phi_L}(x) \mapsto -i\tau^2 \phi_L(T^{-1}(x)) .
\end{eqnarray}

A set of discrete transformations for the gravity variables
consistent with the ones described above for fermion fields
can then be defined as follows:
\begin{eqnarray}
P: \left(\tilde \sigma^{ia}(\vec x,t), A^{\mp}_{ia}(\vec x, t)\right)
&\mapsto&
\left(\tilde\sigma^{ia}(P^{-1}(\vec x, t)), A^{\pm}_{ia}(P^{-1}(\vec
x,t))\right) ,
\nonumber\\
C:\left(\tilde \sigma^{ia}(\vec x, t), A^{\mp}_{ia}(\vec x, t)\right)
&\mapsto& \left(\tilde \sigma^{ia}(\vec x, t), A^{\mp}_{ia}(\vec x,t)
\right) ,
\nonumber\\
T:\left(\tilde \sigma^{ia}(\vec x, t), A^{\mp}_{ia}(\vec x, t)\right)
&\mapsto& \left(\tilde \sigma^{ia}(T^{-1}(\vec x,
t)), A^{\mp}_{ia}(T^{-1}(\vec x, t))\right) .
\end{eqnarray}
These transformations
are consistent with the commutation relations (10).
In dealing with curved spacetime, it is more
convenient to rewrite these transformations in terms of their action on
the one-forms $(e^A, A^{\mp}_a)$. These give
\begin{eqnarray}
P&:& (e^0, e^a ; A^{\mp}_a )
\mapsto (e^0, -e^a ; A^{\pm}_a) ,
\nonumber\\
C&:& (e^A ; A^{\mp}_a )
\mapsto (e^A ; A^{\mp}_a) ,
\nonumber\\
T&:& (e^0, e^a ; A^{\mp}_a )
\mapsto (-e^0, e^a ; A^{\mp}_a) .
\end{eqnarray}

P and T transformations for the Ashtekar variables for pure
 gravity have been discussed
previously \cite{Beng}. However, that discussion did not cover C and CPT, nor
take into account the effect of coupling to fermions. Note also that P and T
are orientation-reversing transformations. While P and C are to be
implemented by unitary transformations, T is to be implemented
anti-unitarily, so that under T, c-numbers are complex conjugated.

We emphasize that the Ashtekar variables are however not necessarily
orientation-reversal invariant. In fact there are
four-manifolds with no orientation reversing
diffeomorphisms. The signature invariant, $\tau$, of a
four-manifold is odd under orientation reversal. Therefore a manifold with
non-vanishing $\tau$ cannot possess an orientation reversing diffeomorphism.
One can show that for the Ashtekar variables, $\int_M
{F^-_a} \wedge F^{-a} - \int_{\overline M} F^+_a \wedge F^{+a} \propto
\tau $ \cite{cps}. Thus for
manifolds with non-vanishing $\tau$'s, $A^+$ can neither be
diffeomorphic nor $SO(3,C)$ gauge equivalent to $A^-$.
%
%
%

To discuss the effect of the discrete transformations in a concise
manner, we first define $A_{AB} = - A_{BA}$ such that
\begin{eqnarray}
A_{0a} &\equiv& {1 \over 2i}(A^{-}_{a} - A^{+}_{a}) ,
\nonumber\\
A_{bc} &\equiv& -{1\over 2}\epsilon^{a}\,_{bc} (A^{-}_{a} + A^{+}_{a}) .
\end{eqnarray}
and keep in mind the effect of the discrete transformations displayed in (30).

The curvature
\begin{equation}
F_{AB} \equiv dA_{AB} + A_{A}\,^{C}\wedge A_{CB}
\end{equation}
then has components
\begin{eqnarray}
F_{0a}&=& {1 \over 2i}(F^{-}_{a} - F^{+}_{a})
\nonumber\\
F_{bc}&=& -{1 \over 2} \epsilon^{a}\,_{bc}(F^{-}_{a}+ F^{+}_{a}) ,
\end{eqnarray}
and the ``torsion" $T_{A}(A^{\mp})$, which depends on $A^{\mp}$, is defined as
\begin{equation}
T^{A} \equiv de^{A}+ A^{A}\,_{B} \wedge e^{B} .
\end{equation}

With all this, it can be shown that the ``chiral" gravitational and
Dirac actions are related to the conventional ones by
\begin{eqnarray}
{\cal S^{\mp}}_{SJS}&=& {1 \over (16\pi{G})}\int_M e^A \wedge e^B
\wedge {\ast}F_{AB}
{\mp}{i \over (16 \pi {G})}\int_M \left[ -d(e^A \wedge T_A) + T^A \wedge T_A
\right]
\nonumber\\
&&-{2\lambda \over 16\pi{G}}\int_M e^0 \wedge e^1 \wedge e^2 \wedge e^3
\end{eqnarray}
and
\begin{equation}
{\cal S^{\mp}_D} = {\cal S_D}+ \int_M \left[\pm{ i\over 4}
\psi^A\epsilon_{ABCD}e^C \wedge e^D \wedge T^B
{\mp}{i\over 2(3!)}d(\psi^A\epsilon_{ABCD}e^B \wedge e^C \wedge e^D)\right],
\end{equation}
where $\psi^A \equiv \phi_L^\dagger \tau^A \phi_L + \chi_L^\dagger \tau^A
\chi_L$.

The combined fermionic and gravitational total action is therefore
\begin{eqnarray}
{\cal S}^{\mp}_{total}&=&{\cal S^{\mp}_D} + {\cal S}^{\mp}_{SJS}
\nonumber\\
&=&{\cal S_D} + {1 \over (16\pi{G})}\int_M e^A \wedge e^B \wedge
{\ast}F_{AB}
\nonumber\\
&&{\mp}{ i \over 2}\int_{M}d\{-{1\over(8\pi G)}e^A\wedge{T_A} +{1 \over
{3!}}(\epsilon_{ABCD}\psi^A e^B \wedge e^C \wedge e^D)\}
\nonumber\\
&&{\mp}{{i}\over{(16 \pi G)}}\int_{M}{\Theta_{A}}\wedge {\Theta}^A
- {2\lambda \over 16\pi{G}}\int_M e^0 \wedge e^1 \wedge e^2 \wedge e^3
\end{eqnarray}
where
\begin{equation}
\Theta_{A} \equiv T_{A} + (2 \pi G)\epsilon_{ABCD} \psi^{B}
e^{C}\wedge e^{D} .
\end{equation}
Observe that the sum of the conventional Einstein-Hilbert-Palatini and
Dirac actions is given by the first two terms in the second line of (37).

We then find that under P (and also CP and CPT), the change is
\begin{eqnarray}
\triangle{{\cal S}^{-}_{D}} &=& P{\cal S}^{-}_{D}P^{-1}
-{\cal S}^{-}_{D}
\nonumber\\
&=&i\int_{M}{1\over{3!}}d(\epsilon_{ABCD}\psi^A e^B \wedge e^C \wedge e^D) -
{1\over 2}\epsilon_{ABCD}{\psi^A}e^C \wedge e^D \wedge T^B
\end{eqnarray}
and the change of the total action is then
\begin{eqnarray}
\triangle {\cal S}^{-}_{total} &=&\triangle {\cal S^{-}_{SJS}}+
\triangle {\cal S}^{-}_{D}
\nonumber\\
&=&i\int_{M}d\{-{1\over (8\pi G)}e^A\wedge{T_A} +{1 \over
{3!}}(\epsilon_{ABCD}\psi^A e^B \wedge e^C \wedge e^D)\}
\nonumber\\
&& + {{i}\over{(8 \pi G)}}\int_{M}{\Theta_{A}}\wedge {\Theta}^A.
\end{eqnarray}
%
%
%
%
%

It should be noted that for spacetimes with Lorentzian signature, the
Ashtekar variables are not real but rather may be required to satisfy reality
conditions which are supposed to be enforced by a suitable inner
product for quantum gravity \cite{ash} . Moreover, the actions are not
explicitly real and their hermiticity could be tied to the inner product
for the yet unavailable quantum theory of gravity. \footnote{Recall that
pure imaginary local Lorentz invariant pieces of the action
are CPT {\it odd}.}

However, if we are
interested in examining second quantized matter in a gravitational
background (for that matter, in flat spacetime), we may enforce the reality
conditions on $A^-$ by hand.  We assume for
our present purposes, that even though we do not at the
moment have an inner product for quantum gravity that will enforce these
reality conditions,
we can pass over to the second order formulation whereby we eliminate the
Ashtekar connections in terms of the vierbein and spinors through the
equations of motion for $A^-$ and enforce the reality conditions by
inspection.

So, varying the total first
order action, ${\cal S}^-_D + {\cal S^-_{SJS}}$, with respect to $A^-$ yields
\begin{equation}
D\Sigma^{-a} = (4\pi {G})\left[ {1\over 3}\epsilon_{bcd}\psi^a e^b \wedge
e^c \wedge e^d + {1\over 2}{\epsilon^a\,_{bc}}\psi^0 e^b \wedge e^c \wedge e_0
+ i\psi^b e^a \wedge e_b \wedge e^0 \right]
\end{equation}
which implies that the Ashtekar connection is
\begin{equation}
A^-_a = i\omega_{0a} - {1\over 2}\epsilon_a\,^{bc}\omega_{bc}
-(2\pi G)\epsilon_{abc}\left[{1\over 2}\epsilon^{bc}\,_{AB}
\psi^A e^B-i\psi^b e^c \right]
\end{equation}
Upon imposing the usual hermiticity requirements on the spinors (which
implies that $\psi^A$ is hermitian) and the requirement that $e^A$ (hence
the spin connection $\omega$) is real, one finds that $\Theta = 0$, which
for the case of pure gravity, reduces to the torsionless condition.
Thus passing over to the second order formulation where the Ashtekar
connection has been eliminated and reality conditions have been imposed,
we find that the change in the action under P (and also CP and CPT) is
\begin{eqnarray}
\triangle {\cal S}^{-}_{total} &=& -{i \over 12}
\int_{M}d\{\epsilon_{ABCD}\psi^A e^B \wedge e^C \wedge e^D\}
\nonumber\\
&=& -{i \over2}\int_{M} \partial_{\mu}[\det(e)j_5^\mu] d^4x
\nonumber\\
&=& -{i\over 2}\int_{M} (\nabla_{\mu} j_5^\mu)\det(e) d^4x
\nonumber\\
&=& -{i\over 2}\triangle Q_5
\end{eqnarray}
where
\begin{eqnarray}
j_5^\mu &=& {\overline \Psi}E^\mu_A \gamma^A\gamma^5\Psi
\nonumber\\
&=& -E^\mu_A(\phi_L^\dagger\tau^A \phi_L - \phi_R^\dagger{\bar
\tau}^A\phi_R)
\nonumber\\
&=& -E^\mu_A(\phi_L^\dagger\tau^A \phi_L + \chi_L^\dagger
\tau^A\chi_L)
\end{eqnarray}
is precisely the chiral current, and $\triangle Q_5$ is the change in the
chiral (axial) charge.
Therefore, it would appear that if the chiral current is not
conserved, the action fails to be {\it hermitian} and invariant under P, CP
and CPT. In this respect, the Adler-Bell-Jackiw anomaly or the axial
anomaly \cite{ABJ}
induced by global instanton
effects may lead to violations of CPT through chirality changing
transitions. Chirality changing transitions due to QCD
instantons have been investigated by others before \cite{rennie},
including 't Hooft \cite{Hooft} in his resolution of the $U(1)_A$ problem.
We shall postpone the discussion on the physical implications and
bounds on the apparent non-hermiticity of the action and the implied
violations of discrete symmetries to a separate report. A related discussion
on the effects of discrete transformations on actions which utilize
self-dual variables in the description of four-dimensional gravity
can be found in Ref. \cite{separep}.

Although the expressions and discussions above are for a single quark or
electron, it is easy to incorporate more matter fields into the
theory simply by replacing $\psi^A$ with
\begin{equation}
\psi^{A} \equiv \sum \Phi_{L}^{\dagger} \tau^{A} \Phi_{L}
\end{equation}
where the sum is over all Weyl fermions, including right-handed fermions
which are to be written as left-handed ones. Conventional Standard Model
Yukawa mass couplings and Higgs fields can be introduced without modifications.
%
%
%
%

\bigskip
\section*{V. The Standard Model with Ashtekar Fields}
\bigskip

We next examine how the quark and lepton fields of the Standard Model
are to be coupled to the Ashtekar fields.  Since we are allowing
couplings to $A^-$ rather than the whole spin connection, only
left-handed fermions can be coupled to the theory.  By writing
all right-handed fields in terms of left-handed ones, it
is possible to couple all the Standard Model quark and
lepton fields to Ashtekar gravity
in a consistent manner.
We shall show in the next section that
this coupling will not disturb the cancellations of all perturbative
anomalies, given the multiplets of the Standard Model.   The
question of how to deal with the global anomaly that may be present
in the theory will be examined in the following section.

We shall label collectively by $W_i T^i$ the gauge connection one-forms
of the Standard Model. The $T^i$'s denote
generators
of the Standard Model gauge group, and there should not be any confusion
between this index i and spatial indices. The covariant derivative in $\cal
S^-_D$ in the action is then replaced by the total covariant derivative
\begin{equation}
{D^-}\phi_{L_I} =  \bigg[(d - iA^-_a{\tau^a \over 2})\delta_{IJ} -i
W_i(T^i)_{IJ}\bigg]\phi_{L_J}
\end{equation}
where the index I associated with $\phi_{L_I}$
denotes internal ``flavor" and/or ``color". In this modified model,
the right-handed Weyl fields of the
conventional Standard Model are to be written as left-handed Weyl
fermions via the relation
\begin{equation}
\phi_{R_I} = -i\tau^2 \chi^*_{L_I}
\end{equation}
So an electron or a single quark of a particular color is represented as a
pair of left-handed Weyl fermions ($\phi_L$ and $\chi_L$), and a left-handed
neutrino is represented by a single left-handed Weyl fermion.

The total action is still to be ${\cal S^{-}_{SJS}} + {\cal S^{-}_{D}}$
with the total covariant derivative of (46) but summed over
all left-handed fermions, including ``right-handed" fermions which are
written as left-handed ones.
Using Eq.(47), right-handed currents can be written
in terms of left-handed currents through
\begin{eqnarray}
J^{{\mu}i}_R &=& \phi^\dagger_{R_I}{\bar \tau}^\mu ({T^i_R})_{IJ}\phi_{R_J}
\nonumber\\
&=& \chi^\dagger_{L_I}{\tau^\mu}({T'^i_L})_{IJ}\chi_{L_J} ,
\end{eqnarray}
with
\begin{equation}
T'^i_L = -(T^i_R)^T .
\end{equation}
Notice however that terms containing the
ordinary gauge connections $W$ are hermitian, so the factor 2 multiplying
terms denoted by (1) and (4) in (22) takes care of the contributions
from terms with $W$ in (2) and (3) which would have been present had we
use the conventional Dirac action. Similarly, this holds also for the
left-handed neutrino
although it is now described by only the term (1) (summed over the
species) instead of (1) and (2). The net result of all this is that
the difference between the ``Modified Standard Model" with gravity
described by ${\cal S^-_D} + {\cal S^-_{SJS}}$ with couplings only to
left-handed fermions and the conventional action is still as
described by (37).
%
%

\bigskip
\section*{VI. Cancellation of Perturbative Gauge Anomalies}
\bigskip

We shall first demonstrate that the
Standard Model defined entirely in terms of left-handed fermions,
and the anti-self-dual Ashtekar fields $A^-$,
is free of perturbative chiral gauge anomalies.

Recall that the anomalies can be determined via Fujikawa's Euclidean
path integral method. We can expand the left-handed multiplet,
${\Psi}_{L_I}
=\left(\matrix{0\cr {\phi}_{L_I}\cr}\right)$, and ${\overline \Psi}_{L_I}$,
in terms of the complete orthonormal set $\{X^L_n, X^R_n\}$ (see, for
instance, Ref. \cite{Fujikawa} ) with
\begin{eqnarray}
X^L_n(x) &=& ({{1-\gamma^5}\over \sqrt 2})X_n(x) , \qquad \lambda_n >0 ,
\nonumber\\
&=& ({{1-\gamma^5}\over 2})X_0(x) , \qquad \lambda_n = 0 .
\nonumber\\
X^R_n(x) &=& ({{1+\gamma^5}\over \sqrt 2})X_n(x) , \qquad \lambda_n >0 ,
\nonumber\\
&=& ({{1+\gamma^5}\over 2})X_0(x) , \qquad \lambda_n = 0 .
\end{eqnarray}
There can be more than one chiral zero modes with their
left-right asymmetry governed by the
Atiyah-Patodi-Singer index theorem. In our present
instance, the curved space Dirac operator contains the full set of $A^\pm$
connections besides $W$, and remains hermitian provided $T_A = 0$; and
$X_n$'s are the eigenvectors with eigenvalues $\lambda_n$ of the full
Euclidean Dirac operator. The chiral
projections in (50) and (51) serve the purposes of defining a chiral
fermion determinant for the effective action as well as selecting only
the $A^-$
Ashtekar fields in the action
$\int (\ast 1){\bar \Psi}_Li\gamma^\mu D_\mu \Psi_L$.
This is achieved by using the Euclidean expansions \cite{alexp}
\begin{eqnarray}
\Psi_L(x) &=& \sum_{ n: \lambda_n \geq 0}a_n X^L_n(x)
\nonumber\\
{\overline \Psi}_L(x) &=& \sum_{n: \lambda_n \geq 0}{\overline b}_n
{X^R_n}^{\dag}(x) .
\end{eqnarray}

The diffeomorphism invariant Euclidean measure for each
left-handed multiplet,
\begin{equation}
\prod_{x} D[{e^{1/2}(x)\overline \Psi}_{L}(x)]D[e^{1/2}(x)\Psi_{L}(x)]
\end{equation}
where $e \equiv det(e)$, is (up to a phase) equivalent to
(see Ref. \cite{Fujikawa})
\begin{equation}
d\mu = \prod_n da_n \prod_n d{\overline b}_n .
\end{equation}

Under a gauge transformation with generator $T^i$,
\begin{equation}
\Psi_L(x) \rightarrow {\it e}^{-i\alpha(x)T^i}\Psi_L(x)
\end{equation}
the measure transforms as
\begin{equation}
d{\mu} \rightarrow
d{\mu}\exp\{-i\int \alpha(x){\cal A}^i(x)dx\}
\end{equation}
The anomaly can be computed as in Ref. \cite{Fujikawa}. This gives
\begin{eqnarray}
{\cal A}^i(x) &=& \sum_{all{\ }n}{X_n}^{\dag}(x)\gamma^5T^iX_n(x)det(e)
\nonumber\\
&=& -{1\over {16{\pi}^2}}Tr\{T^i {1\over 2}\epsilon^{\mu
\nu \alpha \beta}G_{\alpha \beta}G_{\mu\nu}\} ,
\end{eqnarray}
and is proportional to $Tr(T^i\{T^j,T^k\})$. Here, $G^i_{\mu\nu}$ is the
field strength associated with $W^i_\mu$.
So, the condition for cancellation of perturbative gauge anomalies is that
$Tr(T^i\{T^j,T^k\})$ when summed over all fields coupled to $W_iT^i$
vanishes. But as we have seen, if ``right-handed" spinors in the
Standard Model coupled to $W_iT^i_R$
are written as left-handed spinors coupled to $W_iT'^i_L$ such that the
representations $T'_L = -(T_R)^T$, then
\begin{equation}
Tr({T'}_L^i\{{T'}_L^j,{T'}_L^k\}) = -Tr(T^i_R\{T^j_R,T^k_R\}) ,
\end{equation}
and, the condition for the perturbative chiral gauge anomalies of left and
``right"-handed fermions to cancel is
\begin{equation}
Tr(T^i_L\{T^j_L,T^k_L\}) + Tr({T'}_L^i\{{T'}_L^j,{T'}_L^k\}) = 0 ,
\end{equation}
which is precisely equivalent to the well-known condition \cite{Georgi}
\begin{equation}
Tr(T^i_L\{T^j_L,T^k_L\}) = Tr(T^i_R\{T^j_R,T^k_R\}) .
\end{equation}

The Ashtekar fields do not give rise to perturbative anomalies because
the generators of Ashtekar gauge group belong to
su(2), and SU(2) is a ``safe group" where $Tr(T^i\{T^j,T^k\})= 0$ for any
representation.  The introduction of the Ashtekar fields therefore does
not disturb the usual perturbative anomaly cancellation conditions.
But the theory can still be afflicted with global anomalies,
which is the issue we shall address next.

\bigskip
\section*{VII. Global Anomaly and Fermion Content}
\bigskip

Witten \cite{Witten} showed that in four dimensions, theories with an
odd number of Weyl fermion doublets coupled to gauge fields of the
$SU(2)$ group are inconsistent. Without gravity,
the Standard Model has four $SU(2)_{Weak}$ doublets in
each generation, and so it is not troubled by such a global anomaly.
But what about the Ashtekar gauge fields?
Strictly speaking for manifolds of Lorentzian rather than
Euclidean signature, the group is complexified $SU(2)$ and isomorphic to
$SL(2,C)$.  But both these groups have the homotopy group $\Pi_4(G) = Z_2$,
and the non-trivial transformations of the complexified $SU(2)$ group in
$\Pi_4(G)$ are associated with the rotation group.
As Witten \cite{Witten} has argued, the presence of non-trivial elements
of this homotopy group can produce global $SU(2)$ anomalies.
So it would appear that in four dimensions there could be further
constraints on the particle content in order to ensure that the theory be
free of the global anomaly associated with the Ashtekar gauge group.
However, unlike the pure gauge case, gravitational instantons are strongly
correlated with the topology of spacetime, and arguments based on $\Pi_4(G)$
cannot be carried over naively. In what follows, we present a
unified treatment of both gravitational and pure gauge instanton
contributions to the global anomaly for Weyl fermions.

We review the essential points of the global anomaly \cite{Witten, Alwis}.
To begin with, consider a suitable Wick rotation of the
background spacetime into a Riemannian manifold.
Not all manifolds with Lorentzian signature have analytic
continuations to Euclidean signature and vice versa. A more general
setting for this section is the scenario of matter coupled to gravity
in the context of path integral Euclidean Quantum Gravity \cite{eqft}.
The Dirac operator is then hermitian with respect to the
Euclidean inner product $<X|Y> =
\int d^4x \det(e)X^\dagger(x) Y(x)$ \cite{Fujikawa}.
The fermions can then be expanded in terms of the complete set
of the eigenfunctions of the Dirac operator. The expansions will be the
same \cite{alexp} as in Eq.(51) with the eigenfunctions
normalized so that
\begin{equation}
\int_M d^4x \det(e)X^\dagger_m(x)X_n(x) = \delta _{mn}
\end{equation}

Consider next a chiral transformation by $\pi$ which maps each two-component
left-handed Weyl fermion $\Psi_L(x) \mapsto \exp(i\pi\gamma_5)\Psi_L(x)
= \exp(i\pi)\Psi_L(x) = -\Psi_L(x)$. Obviously such a map is a symmetry of
the action. However, the measure is not necessarily invariant under such
a chiral transformation because of the ABJ anomaly. Instead, for each
left-handed fermion, the measure transforms as
\begin{equation}
d\mu \mapsto d\mu \exp(-i\pi \int_M d^4x \det(e)
\sum_n X^\dagger_n(x)\gamma_5 X_n(x))
\end{equation}
The expression
$\int_M d^4x \sum_n \det(e) X^\dagger_n(x)\gamma_5X_n(x)$
is
formally equal to $(n_+ - n_-)$, where $n_{\pm}$ are the number of
normalizable positive and negative chirality zero modes of the Dirac operator.
Upon regularization, the expression works out to be
\begin{eqnarray}
 \sum_n {\det(e)}X^\dagger_n(x){\gamma_5}{X_n(x)}&\equiv&
\lim_{{\cal M} \rightarrow \infty}
  \sum_n \det(e) X^\dagger_n(x){\gamma_5}e^{-(\lambda_n/ {\cal M})^2}X_n(x)
\nonumber\\
&=&\lim_{{\cal M} \rightarrow \infty}\lim_{x' \rightarrow x}Tr\gamma_5 \det(e)
e^{-(i\gamma^\mu{D_\mu}/{\cal M})^2}\sum_n X_n(x)X^\dagger(x')_n
\nonumber\\
&=& -{1\over 384\pi^2}R_{\mu\nu\sigma\tau}\ast R^{\mu \nu\sigma\tau}
\end{eqnarray}

As a result, the index
$(n_+ - n_-) = \sum_n \int_M d^4x \det(e) X^\dagger_n(x)\gamma_5X_n(x) =
-\tau/8$, where $\tau = (1/48\pi^2)\int_M R_{\mu\nu\sigma\tau}\ast
R^{\mu \nu\sigma\tau}$, is the signature invariant of the closed
four-manifold M. (In this report, we restrict our attention to compact
orientable manifolds without boundary. For manifolds with boundaries,
boundary corrections to the Atiyah-Patodi-Singer index theorem must be taken
into account).
Consequently, the change in the measure for {\it each} left-handed Weyl
fermion under a chiral rotation of $\pi$ is precisely
$\exp(i\pi\tau/8)$. If there are altogether $N$ left-handed Weyl fermions in
the theory, the total measure changes by
$\exp(iN \pi\tau/8)$.
But, as emphasized in Ref. \cite{Alwis}, the transformation of $-1$ on the
fermions can also be
considered to be an ordinary $SU(2)$ rotation of $2\pi$. Since $SU(2)$ is a
safe group with no perturbative chiral anomalies (there are no Lorentz
anomalies in four dimensions), the measure must be invariant under all
$SU(2)$ transformations. Thus there is an inconsistency unless this phase
factor, $\exp(iN \pi\tau/8)$, is always unity.

It is known that for consistency of parallel
transport of spinors for {\it topological four-manifolds}, $\tau$ must be a
multiple of 8 for spin structures to exist \cite{EGH}.  This result also
follows from $n_+ - n_- = -\tau/8$, since the index must always be an
integer. However, a theorem due to Rohlin \cite{Rohlin} states the
signature invariant of a {\it smooth} simply-connected
closed spin four-manifold is divisible by 16. If one restricts to
four-manifolds with signature invariants which are multiples of 16
in quantum gravity, or in semiclassical quantum field theory with
left-handed spinors, one
allows for these four-manifold backgrounds only, then under a chiral
rotation of $\pi$ the measure is invariant regardless of $N$. Otherwise,
if we allow for all manifolds with signature invariants which are
multiples of 8, consistency with the global anomaly cancellation
requires that $N$ must be even. In either instance, the net index summed
over all fermion fields is even.

It is not presently clear what the integration range for $A^-$ should
be.  But, it is quite likely that it should extend over a wider
class of manifolds than those which admit ordinary spin
structures \cite{hp}.
The considerations outlined above suggest that in order to do this,
we would have to allow for couplings among the various fermion fields,
for example via Yukawa couplings to spin-0 fields, and particularly via
couplings to gauge fields, which is what occurs in grand unification
theories(GUTs). The reason can be stated as follows.
When $\tau$ is not a multiple of 8, it is not possible to lift the
$SO(4)$ bundle to its double cover $Spin(4)$ bundle since the second
Stiefel-Whitney class, $w_2$, is non-trivial. However, given a general grand
unification (simple) simply-connected gauge group $G$ with a $Z_2 = \{e, c\}$
in its center, it is possible to construct
{\it generalized spin structures} with gauge group $Spin_G(4) =
{\{Spin(4)\times G\}/{Z_2}}$ where the $Z_2$ equivalence relation
is defined by $(x,g) \equiv (-x,cg)$ for
all $(x,g) \in Spin(4)\times G$ \cite{hp}. Note that $Spin_G(4)$ is
the double cover of $SO(4)\times (G/Z_2)$.
Such spinors then transform according to double
cover $Spin_G(4)$ group, and the parallel transport of fermions
does not give rise to any inconsistency. However, the
additional quantum field theoretic global anomaly cancellation condition
must still be satisfied if the theory is to be consistent.

Now, in a scenario where all of the fermions
are coupled to one another via gauge fields,
the index theorem should be applied only to the trace current
involving the sum over all the fermion fields. The resultant restriction on
$N$ is the condition that the index for the {\it total}
Dirac operator with $Spin_G(4)$ connections, $ N_+ - N_-$, is even.
Otherwise, Green's functions which contain
$(N_+ - N_-)$ more ${\overline \Psi}_L$ than $\Psi_L$ variables in the odd
index sector will be inconsistent.
The relevant index of the total Dirac operator in the presence of an
additional grand unification gauge field besides the spin connection is
\begin{equation}
N_+ - N_- = -N{\tau \over 8} - {1\over{8\pi^2}}\int_M Tr(F\wedge F)
\end{equation}
where $F = F^i{\cal T}_i$ is the curvature of the internal grand
unification gauge connection.
The global anomaly consistency condition given above then
implies that for arbitrary $\tau$'s there can only be $16k$
fermions in total in the theory, where k is an integer; and the gauge
group should be selected such that the instanton number,
${1\over{8\pi^2}}\int_M Tr(F \wedge F)$, is always even for non-singular
configurations.

If one counts the number of left-handed Weyl fermions that are
coupled to the Ashtekar connection for the Standard Model,
one finds that the number is 15 per generation, giving a
total of 45 for 3 generations.
This comes about because each bispinor is coupled twice to the Ashtekar
connection while each Weyl spinor is coupled once (e.g. for the first
generation, the number is 2 for each electron and each up or down quark of a
particular color, and 1 for each left-handed neutrino.)
Thus even if one restricts to $\tau =8k$, the global anomaly with respect
to the Ashtekar gauge group seems to
imply that there should be additional particle(s).   For example,
there could be a partner
for each neutrino, making $N$ to be 16 per generation, or a partner for
just the $\tau$-neutrino, or even four generations.
As a result, even with $\tau = 8k$, grand unification schemes based upon
groups such as $SU(5)$ and odd number of Weyl fermions would be inconsistent
when coupled to gravity.

The net conclusion is therefore that inclusion of manifolds with arbitrary
$\tau$'s in the integration range is possible, provided we allow
for generalized spin structures, with a total of $N = 16k$ fermions,
in order that the global anomaly is absent.
In the event that the structures
are defined by simple GUT groups, the simplest choice would be
$SO(10)$.  It is easy to check that the 16 Weyl fermions in
the $SO(10)= Spin(10)/{Z_2}$ GUT indeed belong to the 16-dimensional
representation of $Spin(10)$ and satisfy the generalized spin structure
equivalence relation for $\{Spin(4)\times Spin(10)\}/{Z_2}$.

It is worth emphasizing
that this generalized spin structure implies an additional
``isospin-spin'' relation in that
fermions must belong to $Spin(10)$
representations, while bosons must belong to $SO(10)$ representations
of the GUT. This has implications for spontaneous symmetry breaking via
fundamental bosonic Higgs, which cannot belong to the spinorial
representations of $SO(10)$ if one allows for arbitrary $\tau$'s.
More generally,  the presence of the extra particle(s) implied by
global anomaly considerations can generate masses for neutrinos,
thereby giving rise to neutrino
oscillations, and may also play a significant role in the cosmological
issue of  ``dark matter".

A more mathematical treatment of the perturbative and global anomaly
cancellation conditions we have discussed can be given.
It is based on the fact that
the chiral fermion determinant is actually a {\it section} of the
determinant line bundle \cite{APS, Moore, Bismut, Freed}. In order for the
functional integral over the remaining fields to make sense, there must
not be any obstruction to the existence of a global section. It is
therefore necessary for the first Chern class of the determinant line
bundle to vanish \cite{APS, Moore}. This gives the result for
perturbative anomaly
cancellation. However, a vanishing first Chern class does not
necessarily imply a trivial connection for the determinant line bundle.
There can be a further obstruction to a consistent definition of the
chiral fermion determinant due to non-trivial parallel transport
around a closed loop. Indeed, it can be shown that under
suitable assumptions, the holonomy of the Quillen-Bismut-Freed connection
of the determinant line bundle is given by
$(-1)^{index (Dirac(M,A))}exp(-2i\pi\xi)$.
Here $\xi = {1 \over 2}(\eta + h)$, with $\eta$ and $h$
being the spectral asymmetry and number of zero modes respectively, of the
Dirac operator on the five-dimensional manifold $M\times S^1$
endowed with a suitably scaled metric on $S^1$ \cite{Bismut, Freed}.
In order to evaluate $\xi$,
one can view $M\times S^1$ as the boundary of a six-dimensional
manifold, $M_6$, with a product metric on the boundary, and apply the
Atiyah-Patodi-Singer index theorem to obtain
\begin{equation}
index (Dirac)_{M_6} = -\int_{M_6}
{{\hat A}}\wedge{ch(F)}-\xi(M\times{S^1}) .
\end{equation}
It is instructive to work out the terms and their implications explicitly.
For the case relevant to us, $M$ is four-dimensional and we pick out the
six-forms in the integrand in the first term on the RHS. These are of the
form $Tr({F}\wedge{F}\wedge{F})$ and $R_{AB}\wedge R^{AB}\wedge{Tr(F)}$.
The former is the abelian anomaly in D+2 dimensions (here D=4) which
gives rise to the perturbative non-abelian gauge anomaly in
D-dimensions \cite{APS, Moore, WZW}. It vanishes when
the perturbative anomaly-free condition $Tr({\cal T}_i \{{\cal T}_j, {\cal
T}_k\}) = 0$ is satisfied.
The latter, which is the gravitational-gauge cross term, is required to
vanish $(Tr(F) = 0)$ for the absence of mixed
Lorentz-gauge anomalies when there are couplings to abelian fields
\cite{Nieh}. These terms occur in the holonomy theorem of Bismut and Freed
because the holonomy measures the non-triviality of the
connection of the determinant line bundle, and pertubative anomalies which
correspond to the non-triviality of the first Chern class must
necessarily contribute. In this spirit, discussions of global anomalies make
sense only if perturbative anomalies cancel.
With the perturbative anomaly cancellation
conditions, we therefore conclude that $\xi$ is an integer
since $index (Dirac)_{M_6}$ is an integer. Thus the holomony theorem
tells us that the remaining obstruction i.e. the global anomaly is due to
$(-1)^{index (Dirac(M, A))}$, and the consistency condition is again the
requirement that the index of the total chiral Dirac operator in four
dimensions with generalized spin structure is even.

We have presented our arguments of the global anomaly in terms of the
Ashtekar variables for definiteness. We would like
to end by emphasizing that our results would also obtain
in the setting of Weyl fermions coupled to conventional gravity and
ordinary spin connections instead of the Ashtekar connections \cite{cfg}.
This is essentially because the anomaly can be understood as coming from the
inconsistent change in the fermionic measure or the chiral
fermion determinant, and for both schemes, the change is $\exp[-i\pi(N_+
- N_-)] = (-1)^{index (Dirac)}$.

\bigskip\bigskip\bigskip
\section*{Acknowledgments}
\bigskip

The research for this work has been supported by the Department of Energy
under Grant No. DE-FG05-92ER40709, the NSF under Grant No. PHY-9396246,
and research funds provided by the Pennsylvania State University.
One of us (CS) would like to thank Abhay Ashtekar, Lee Smolin, Jose Mourao
and other members of the
Center for Gravitational Physics and Geometry for encouragement and helpful
discussions. LNC would like to thank Peter Haskell for a helpful discussion.


\begin{thebibliography}{99}
\bibitem{rema} From Eq. (42), it can be seen that in the second
order formalism, the Ashtekar-Sen connection contributes a cubic
fermionic term to the Dirac equation when we vary with respect to the
spinors. This cubic term can be cancelled by adding a
quartic term proportional to $G \int_M (\ast 1)\psi^A \psi_A$ to the action.
However, this extra term in the action is separately invariant under C, P
and T, and hence do not affect our discussions on the discrete symmetries
later on.
\bibitem{alexp} An alternative scheme is to expand
$\Psi_L$ and ${\overline \Psi}_L$ in terms of $X_n$ and
${\tilde X}^\dagger_n$, where $X_n$ and ${\tilde X}_n$ are
eigenfunctions of the positive semidefinite hermitian operators
${\cal D}^\dagger{\cal D}$ and ${\cal D}{\cal D}^\dagger$ respectively.
Here ${\cal D}$ is the total chiral
Dirac operator ${\tau^A}{E_A}_\lfloor(id + A^- + W)$. See, for instance, Refs.
\cite{Fujikawa} and \cite{APS}.
\end{thebibliography}
\end{document}